\documentclass[preprint,12pt]{elsarticle}

\usepackage{amssymb}
\usepackage{amsmath}
\usepackage{bm}
\usepackage{subfig}
\biboptions{sort&compress}
\usepackage{floatrow}
\usepackage{booktabs}
\journal{arXiv}

\begin{document}

\begin{frontmatter}



\title{Precise Damage Shaping in Self-Sensing Composites Using Electrical Impedance Tomography and Genetic Algorithms}

\author{Hashim Hassan}
\author{Tyler N. Tallman}
\address{School of Aeronautics and Astronautics, Purdue University, West Lafayette, IN, 47907, United States}

\begin{abstract}
Fiber-reinforced composites with nanofiller-modified polymer matrices have immense potential to improve the safety of high-risk engineering structures. These materials are intrinsically self-sensing because their electrical conductivity is affected by deformations and damage. This property, known as piezoresistivity, has been extensively leveraged for conductivity-based damage detection via electrical resistance change methods and tomographic imaging techniques such as electrical impedance tomography (EIT). Although these techniques are very effective at detecting the presence of damage, they suffer from an inability to provide precise information about damage shape, size, or mechanism. This is particularly detrimental for laminated composites which can suffer from complex failure modes, such as delaminations, that are difficult to detect. To that end, we herein propose a new technique for precisely determining damage shape and size in self-sensing composites. Our technique makes use of a genetic algorithm (GA) integrated with realistic physics-based damage models to recover precise damage shape from conductivity changes imaged via EIT. We experimentally validate this technique on carbon nanofiber (CNF)-modified glass fiber-reinforced polymer (GFRP) laminates by considering two specific damage mechanisms: through-holes and delaminations. Our results show that this novel technique can accurately reconstruct multiple through-holes with radii as small as 1.19 mm and delaminations caused by low velocity impacts. These findings illustrate that coupling piezoresistivity with conductivity-based spatial imaging techniques and physics-based inversion strategies can enable damage shaping capabilities in self-sensing composite structures. 
\end{abstract}


\end{frontmatter}

\section{Introduction}
Fiber-reinforced composites are ubiquitous in aerospace, automotive, mechanical, and light-weight civil structures due to their high strength-to-weight ratios, good fatigue performance, and design flexibility. However, it is well known that composite structures are susceptible to matrix cracking, delamination, and fiber breakage. If left unchecked, these damage modes can accumulate and cause catastrophic failure. For structures in safety-critical applications such as aircraft and bridges, this may result in significant loss of life. Therefore, damage must be inspected and preventive measures taken before it reaches a critical level. 

Traditional damage tolerant design approaches that rely on intermittent inspections via nondestructive evaluation (NDE) often prove insufficient for composites due to the complexity of their damage initiation and growth mechanisms. Structural health monitoring (SHM) \citep{sohn2003review, lynch2007overview} is a promising complement to traditional NDE as it allows for continuous, real-time monitoring of damage initiation and growth in composite structures. Conventional SHM methodologies rely on the use of external sensors to monitor the condition of the structure. Three common categorizations of conventional SHM are vibration-based SHM, guided wave-based SHM, and embedded sensor-based SHM. 

Vibration-based SHM \citep{carden2004vibration, montalvao2006review} monitors changes in the modal parameters of the structure such as natural frequencies, mode shapes, and curvature changes. These methods have been used to detect delaminations \cite{zou2000vibration, valdes1999delamination} and matrix cracks \cite{kessler2002damage} in composite plates and also in circular cylindrical composite shells \cite{ip2002locating}. Guided wave-based SHM \cite{mitra2016guided, raghavan2007guided, croxford2007strategies} uses ultrasonic wave propagation to detect the presence of damage. The waves are dispersed or attenuated when they encounter a flaw or a crack. The most common types of waves used for detecting cracks and delaminations include Lamb waves \cite{su2006guided, paget2003damage} and Rayleigh waves \cite{chakrapani2014interaction, hughes2019comparative}. Embedded sensor-based SHM uses discrete sensors integrated within the structure to monitor its condition. The most common types of sensors used for composites are strain gauges \cite{ajovalasit2011advances, zike2014correction, zhou2017strain}, fiber-optic sensors \cite{leung2005delamination, lee2018distributed, ramakrishnan2016overview, leng2003structural}, and piezoelectric sensors \cite{giurgiutiu2011structural, kalhori2017inverse}.

Although conventional SHM methodologies are more widely used, they have certain limitations. Vibration-based SHM, for example, is often not sensitive enough to detect small-sized defects. Both guided wave-based and embedded sensor-based SHM require physical integration of the sensors and actuators within the structure which can sometimes weaken the material. Another overarching limitation is that these methods only provide localized, discrete, and often limited information about the condition of the structure. Consequently, recent research focus has shifted toward the development of composites with intrinsic self-sensing properties to alleviate the limitations of conventional SHM.

Self-sensing composites possess one or more properties that allow them to `sense' changes to their mechanical, thermal, or chemical state. That is, the material itself functions as a sensor. In this area, nanofiller-modified composites have received considerable attention \cite{li2019review, costa2017high, stassi2014flexible, groo2020laser, groo2021fatigue}. These materials exhibit changes in their electrical behavior due to deformations and damage. This self-sensing property, known as piezoresistivity, has been extensively studied for SHM in combination with conductivity-based spatial imaging techniques such as electrical impedance tomography (EIT) because this combination allows for the mapping of mechanical effects through conductivity changes.

Beyond just allowing for spatial mapping of mechanics in self-sensing composites, EIT has recently emerged as a potential SHM modality because it is low-cost, non-invasive, and has nearly real-time imaging capabilities. Broadly speaking, and as summarized in \citep{tallman2020structural}, EIT has been used for SHM in intrinsically and extrinsically self-sensing composites. Intrinsically self-sensing composites incorporate conductive nanofillers in the polymer matrix. For such materials, EIT has been used for detecting through-holes \cite{tallman2015damage}, impact damage \cite{gallo2016spatial}, matrix cracking \cite{dai2016novel}, delaminations \cite{haingartner2020improved}, and UV radiation damage \cite{rastogi2021structural, clausi2019direct} in composites incorporating carbon nanotubes (CNTs) \cite{dai2016novel}, carbon nanofibers (CNFs) \cite{tallman2017inverse, tallman2014damage, hassan2020failure,hassan2020comparison}, and carbon black (CB) \cite{tallman2015damage}. Although most research has focused on thin plates, there has been some work in using EIT for damage detection in non-planar structures as well \cite{thomas2019damage, jauhiainen2021nonplanar}. Extrinsically self-sensing composites do not incorporate nanofillers but make use of externally applied sensing media. In this area, EIT has been used for detecting strain and damage in composites using nanofiller-based thin films \cite{hou2007spatial, hou2007electrical} and paints \cite{lestari2016sensing, loyola2013spatial}. Aside from composites, EIT has been used extensively for detecting damage due to holes \cite{gupta2017self, nayak2019spatial}, cracks \cite{smyl2018detection}, and moisture flows \cite{hallaji2015electrical} in self-sensing cementitious materials \cite{hallaji2014electrical, hallaji2014new}.

From the preceding discussion, it is clear that EIT can effectively detect and localize numerous damage modes in self-sensing materials. However, a critical limitation is that it provides little-to-no information about precise damage shape, size, or mechanism. This information is crucial for accurate prognostication of structural health. For example, if the precise size and location of a delamination can be ascertained, structural failure can be predicted and averted. Herein, we recognize the opportunity to address this limitation and advance the state-of-the-art in SHM of self-sensing composites. We propose a novel methodology for precisely determining damage shape and size from electrical measurements in self-sensing composites. Our unique approach uses a genetic algorithm (GA) to inversely compute damage geometry based on imaged conductivity changes and boundary voltage measurements. We experimentally investigate the potential of this approach on two damage mechanisms---through-holes and delaminations---in CNF-modified GFRP laminates.

The remainder of this manuscript is organized as follows. First, we mathematically formulate the EIT forward and inverse problems. Second, we present and formulate the GA-enabled damage shaping problem. Third, we describe the composite manufacturing procedures. Fourth, we discuss the electrical, through-hole, and impact testing procedures. Fifth, we present the EIT results along with the results from our GA-enabled damage shaping methodology and discuss their significance. Finally, we end with a brief summary, conclusions, and possible paths for future work. 

\section{Electrical Impedance Tomography}
\subsection{Forward Problem}
\begin{figure}
	\centering
	\includegraphics[width=0.75\textwidth]{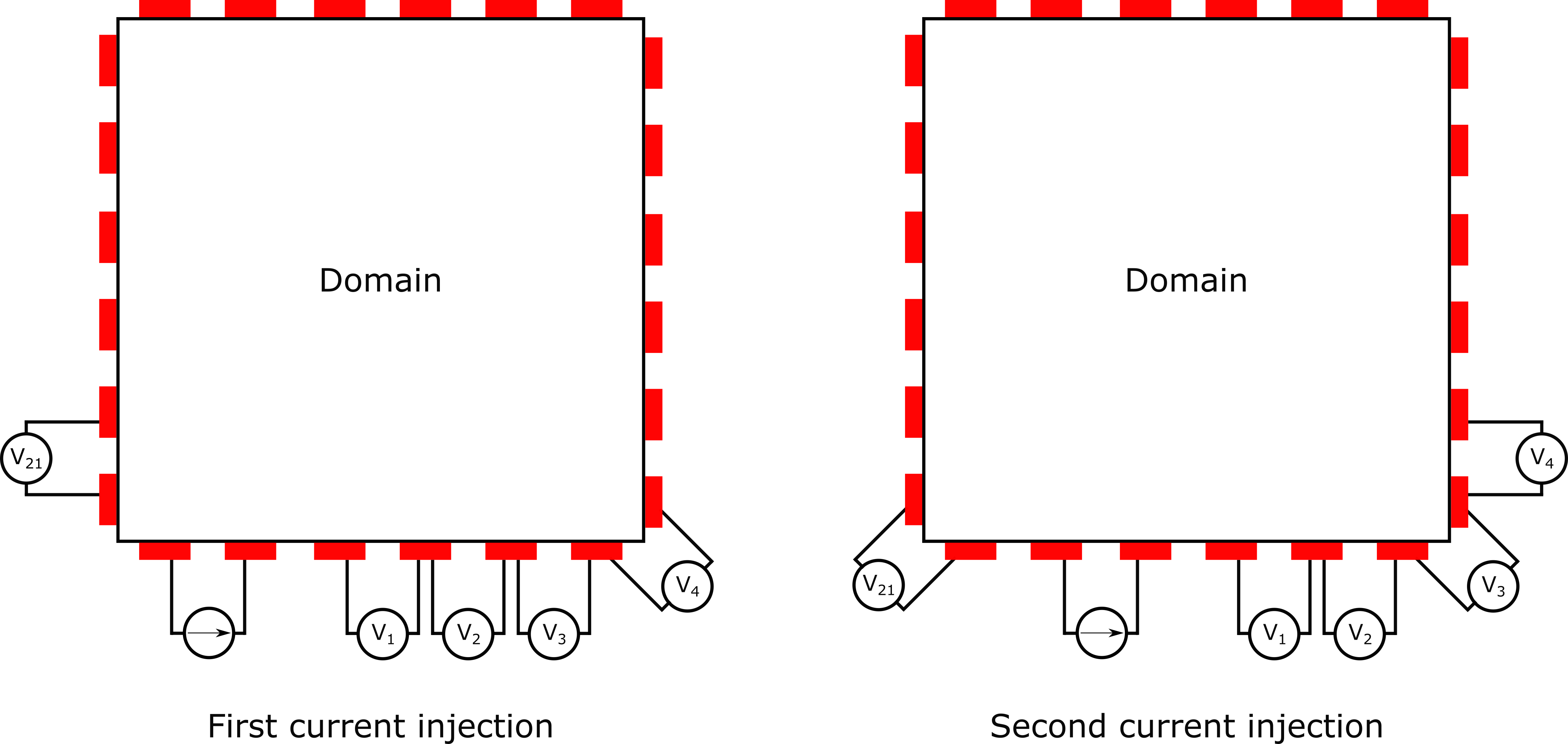}
	\caption{Illustration of first and second current injection and voltage measurement schemes. The periphery of the domain is instrumented with electrodes, indicated by the red rectangles. Current is injected between the first pair of electrodes and voltage differences are measured between the remaining electrode pairs. The current injection is then moved to the next electrode pair and voltage differences are again measured. This process repeats until all adjacent electrode pairs have received a current injection.}
	\label{eit-injection}
\end{figure}

As discussed earlier, EIT is a non-invasive method of imaging the internal conductivity distribution of a domain. EIT works by minimizing the difference between a vector of experimentally measured voltages and a vector of voltages computed numerically. The experimental voltage collection procedure is illustrated in Figure \ref{eit-injection}. The domain to be imaged is lined with electrodes. Current is injected between the first pair of electrodes and voltage differences are measured between the remaining electrode pairs. The current injection is then moved to the next electrode pair and voltage differences are again measured between the remaining electrode pairs. This process is repeated until all electrode pairs have received one current injection and a vector of \(L(L-3)\) voltage differences has been obtained, where \(L\) is the total number of electrodes.

The numerical voltages are computed by solving the EIT forward problem. To formulate the forward problem, consider Laplace's equation for steady-state diffusion in the absence of internal current sources. This is shown in equation (\ref{eq1}), where \(\sigma\) is the internal conductivity of the domain and \(\phi\) is the domain potential. We enforce two complete electrode model boundary conditions on equation (\ref{eq1}). The first boundary condition, shown in equation (\ref{eq2}), assumes a voltage drop across the electrodes due to the contact impedance between the domain and the electrodes. The second boundary condition, shown in equation (\ref{eq3}), enforces conservation of charge. In these equations, \(z_l\) is the contact impedance between the \(l\)th electrode and the domain, \(\boldsymbol{n}\) is an outward pointing normal vector, \(E_l\) is the length of the \(l\)th electrode, and \(V_l\) is the voltage of the \(l\)th electrode. 

\begin{equation}\label{eq1}
    \nabla\cdot\sigma\nabla\phi=0
\end{equation}

\begin{equation}\label{eq2}
    \phi + z_l\sigma\nabla\phi\cdot\boldsymbol{n}=V_l
\end{equation}

\begin{equation}\label{eq3}
    \sum_{l=1}^{L}\int_{E_l}\sigma\nabla\phi\cdot\boldsymbol{n} \hspace{3pt} \mathrm{d}S_l=0
\end{equation}

Equations (\ref{eq1}) to (\ref{eq3}) are often solved via the finite element (FE) method, as shown in equation (\ref{eq4}). In this equation, \(\boldsymbol{\Phi}\) is a vector of domain potentials, \(\boldsymbol{V}\) is a vector of electrode voltages and \(\boldsymbol{I}\) is a vector of injected currents. The matrices \(\boldsymbol{A_M}\), \(\boldsymbol{A_Z}\), \(\boldsymbol{A_W}\), and \(\boldsymbol{A_D}\) are formed as shown in equations (\ref{eq5}) to (\ref{eq8}), where \(N_i\) is the \(i\)th FE basis function. For an explicit solution to these equations for linear elements, interested readers are directed to \citep{tallman2020structural}.

\begin{equation}\label{eq4}
\begin{bmatrix}
\boldsymbol{A_M} + \boldsymbol{A_Z} & \boldsymbol{A_W} \\
\boldsymbol{A_W^T} & \boldsymbol{A_D}\\
\end{bmatrix}
\begin{bmatrix}
\boldsymbol{\Phi}\\
\boldsymbol{V}
\end{bmatrix}=
\begin{bmatrix}
\boldsymbol{0}\\
\boldsymbol{I}
\end{bmatrix}
\end{equation}

\begin{equation}\label{eq5}
    A^e_{M \hspace{3pt} ij} = \int_{\Omega_e}\frac{\partial N_i}{\partial x_k} \sigma_{kl} \frac{\partial N_j}{\partial x_l} \hspace{3pt} \mathrm{d}\Omega_e
\end{equation}

\begin{equation}\label{eq6}
    A_{Z\hspace{3pt} ij} = \sum_{l=1}^L \int_{E_l}\frac{1}{z_l} N_i N_j \hspace{3pt} \mathrm{d}S_l
\end{equation}

\begin{equation}\label{eq7}
    A_{W \hspace{3pt} li} = -\int_{E_l}\frac{1}{z_l} N_i \hspace{3pt} \mathrm{d}S_l
\end{equation}

\begin{equation}\label{eq8}
    A_D = \text{diag}\bigg(\frac{E_l}{z_l}\bigg)
\end{equation}

\subsection{Inverse Problem}
The goal of the EIT inverse problem is to recover the conductivity distribution of the domain interior using the measured and numerically computed boundary voltages. Herein, we will make use of a one-step linearization scheme to solve the inverse problem. This involves collecting a set of pre-damage voltages, a set of post-damage voltages, and then finding the conductivity change distribution that minimizes the difference between the two sets of voltages. We begin by defining the vector \(\delta\boldsymbol{V}\) as the difference between voltages collected at times \(t_2\) and \(t_1\), as shown in equation (\ref{eq9}).

\begin{equation}\label{eq9}
    \delta\boldsymbol{V} = \boldsymbol{V}(\sigma_2,t_2) - \boldsymbol{V}(\sigma_1,t_1)
\end{equation}

Next, we define an equivalent numerical vector, \(\boldsymbol{W}(\delta\boldsymbol{\sigma})\), as shown in equation (\ref{eq10}). Here, \(\boldsymbol{F}(\cdot)\) represent the voltages computed by solving the forward problem at the conductivity in the argument, \(\boldsymbol{\sigma_0}\) is the baseline or undamaged conductivity, and \(\delta\boldsymbol{\sigma}\) is the conductivity change vector we seek. The boldfaced symbols indicate vector quantities that have been discretized via finite elements.

\begin{equation}\label{eq10}
    \boldsymbol{W}(\delta\boldsymbol{\sigma}) = \boldsymbol{F}(\boldsymbol{\sigma_0} + \delta\boldsymbol{\sigma}) - \boldsymbol{F}(\boldsymbol{\sigma_0})
\end{equation}

We then linearize the first term on the right-hand side of equation (\ref{eq10}) using a Taylor series expansion about \(\boldsymbol{\sigma_0}\) and retain only the linear terms, as shown in equation (\ref{eq11}).

\begin{equation}\label{eq11}
    \boldsymbol{F}(\boldsymbol{\sigma_0} + \delta\boldsymbol{\sigma}) \approx \boldsymbol{F}(\boldsymbol{\sigma_0}) + \frac{\partial \boldsymbol{F}(\boldsymbol{\sigma_0})}{\partial \boldsymbol{\sigma}}\delta\boldsymbol{\sigma}
\end{equation}

Substituting equation (\ref{eq11}) into equation (\ref{eq10}) and defining \(\boldsymbol{J}\) = \(\partial \boldsymbol{F}(\boldsymbol{\sigma_0})/\partial\boldsymbol{\sigma}\) as the sensitivity matrix yields equation (\ref{eq12}). The one-step linearized inverse problem then seeks the conductivity change, \(\delta\boldsymbol{\sigma}^*\), that minimizes the difference between \(\boldsymbol{W}(\delta\boldsymbol{\sigma})\) and \(\delta\boldsymbol{V}\) in the constrained linear least squares sense, as shown in equation (\ref{eq13}).

\begin{equation}\label{eq12}
    \boldsymbol{W}(\delta\boldsymbol{\sigma}) \approx \boldsymbol{J}\delta\boldsymbol{\sigma}
\end{equation}

\begin{equation}\label{eq13}
    \delta\boldsymbol{\sigma}^* = \underset{\boldsymbol{-\sigma_0}\leq\delta\boldsymbol{\sigma}\leq0.01\boldsymbol{\sigma_0}}{\text{min}}\frac{1}{2}\Bigg(\Bigg|\Bigg|\begin{bmatrix}
    \boldsymbol{J}\\
    \alpha\boldsymbol{L}\\
    \end{bmatrix}\delta\boldsymbol{\sigma} - \begin{bmatrix}
    \delta\boldsymbol{V}\\
    \boldsymbol{0}\\
    \end{bmatrix}\Bigg|\Bigg|^2_2\Bigg)
\end{equation}

A regularization term, \(\boldsymbol{L}\), has been included in equation (\ref{eq13}) due to the ill-posed nature of the EIT inverse problem. Here, we will use the discrete Laplace operator as the regularization term. The amount of regularization is controlled by the scalar parameter \(\alpha\). Also, note that the search for \(\delta\boldsymbol{\sigma}^*\) is constrained. The constraints used here are based on realistic assumptions about the conductivity change. It is well understood that damage such as cracks, through-holes, and delaminations manifests as a decrease in conductivity. The maximum decrease that is physically possible has a magnitude equal to the baseline conductivity and we expect the maximum increase to be relatively small. Therefore, we use 100\% change for the lower bound and 1\% change for the upper bound. This is reflected in the limits shown in equation (\ref{eq13}).

\section{GA-Enabled Damage Shaping Problem}
As discussed earlier, EIT has been successfully used to detect various kinds of damage in self-sensing composites. However, due to the ill-posed nature of the inverse problem, the spatial resolution of EIT is somewhat indistinct inasmuch as precise geometric features of conductivity artifacts cannot be sharply reconstructed. Considerable research has aimed at developing methodologies for precise artifact shape reconstruction from EIT images \cite{liu2020shape, liu2020bool, fan2021new}. While these approaches are undoubtedly powerful, they are unaware of the underlying physics of the material and consequently cannot discern between various damage modes. Therefore, in order to accurately determine damage shape and size from EIT-imaged conductivity changes, specific damage mechanisms must be integrated with suitable inversion strategies. 

The goal of the damage shaping problem presented here is to recover the geometry of specific damage modes from EIT-imaged conductivity changes and boundary voltages by building some knowledge about the damage physics into the shaping algorithm. However, recovering mechanics from electrical measurements is an ill-posed, multi-modal inverse problem. This is because there is no well-defined mathematical relationship between damage geometry and conductivity change. In the absence of a suitable inversion algorithm and realistic constraints on the search space, the search may converge to a locally optimum, non-physical solution. In order to address this problem we need a global search algorithm and constraints based on realistic damage physics.

Genetic algorithms (GAs) \cite{golberg1989genetic, mitchell1998introduction} are a family of global search algorithms inspired by natural evolution. GAs are widely used for optimizing multi-modal functions by generating and evolving a population of solutions dispersed inside a search space. The population of solutions evolves through a process analogous to natural selection. Herein, we do not aim to develop a new algorithm but to instead use a well-established GA to solve the multi-modal damage shaping problem. To that end, we will make use of a GA developed by WA Crossley and used by Raghavan et al. \cite{raghavan2008spectral} for jet engine turbine blade NDE.


In addition to a global search algorithm, we require geometric models that realistically describe how different damage modes affect material conductivity. These models will need to be integrated with the GA so that only physically viable solutions are generated and selected. We can now formulate the GA-enabled damage shaping problem. We begin by defining the vector \(\delta\boldsymbol{F}(s_{GA})\) as shown in equation (\ref{eq14}), where \(\boldsymbol{F}(\boldsymbol{\sigma}(s_{GA}))\) are the voltages predicted by solving the EIT forward problem for a domain containing damage described by the GA-generated parameter, \(s_{GA}\). This parameter describes the shape and location of a specific damage mode based on a geometric damage model.

\begin{equation}\label{eq14}
    \delta\boldsymbol{F}(s_{GA}) = \boldsymbol{F}(\boldsymbol{\sigma}(s_{GA})) - \boldsymbol{F}(\boldsymbol{\sigma_0})
\end{equation}

The goal of the damage shaping problem then is to seek the optimum shape parameter, \(s_{GA}^*\), that minimizes the \(l_1\)-norm of the difference between \(\delta\boldsymbol{F}(s_{GA})\), given by equation (\ref{eq14}), and \(\delta\boldsymbol{V}\), given by equation (\ref{eq9}). This can be cast as the optimization problem shown in equation (\ref{eq15}). In this work, we will develop geometric models for two damage modes---through-holes and delaminations---and integrate them with the GA to solve equation (\ref{eq15}). These models are described later in \S\ref{results_and_discuss}.

\begin{equation}\label{eq15}
    s_{GA}^* = \underset{s_{\text{min}} \leq s_{GA}\leq s_{\text{max}}}{\arg\min} (\underbrace{|\delta\boldsymbol{V} - \delta\boldsymbol{F}(s_{GA})|_1}_{f})
\end{equation}

In the above equation, the parenthetical term is the fitness function, \(f\). During a single search, as the population of solutions evolves, \(f\) decreases and finally attains a stable value. This occurs when the candidates in the population have achieved a certain level of genetic similarity. This is quantified as the bit-string affinity (BSA) and is a pre-specified value. When the BSA is achieved, further evolution ceases. The BSA is one of two convergence criteria that we will use herein. The second convergence criterion is the maximum number of generations. That is, after the population has evolved a pre-specified number of times, further evolution ceases. At this point the optimum solution, \(s^*_{GA}\), and minimum fitness function value, \(f^*\), are recorded and a second search initiates using updated values of \(s_{\text{min}}\) and \(s_{\text{max}}\). This is repeated until the following global convergence criterion is satisfied, where \(f_n^*\) is the minimum fitness function value obtained after the \(n\)th search.

\begin{equation}
    |f_n^* - f_{n+1}^*| \leq 1\times10^{-3}
\end{equation}

At this point, no further searches are performed using the GA and the final solution attained is treated as the converged solution.

\section{Composite Manufacturing}
To experimentally validate the proposed GA-enabled damage shaping methodology, GFRP laminates with 1.0 wt.\% CNF-modified polymer matrices were manufactured based on the procedure described by Tallman et al. \cite{tallman2014damage}. The CNF-modified matrix was prepared by mixing epoxy resin (Fibre Glast 2000) with the appropriate amounts of Pyrograf III PR-24-XT-HHT CNFs (Applied Sciences), surfactant (Triton X-100), and acetone. Acetone aids mixing by lowering the viscosity and the surfactant facilitates dispersion of the CNFs by modifying the surface chemistry of the mixture. An epoxy-to-acetone volume ratio of 2:1 and a surfactant-to-CNF weight ratio of 0.76:1 were used. The modified epoxy was then mixed in a planetary centrifuge for 3 minutes and sonicated in a bath sonicator for 1 hour for every 10 g of the mixture. The bath sonicator operated at a power of 35 W and a frequency of 45 kHz. After sonication, the mixture was stirred on a magnetic hot plate stirrer for 24 hours at a temperature of 60 $^{\circ}$C and a stirring speed of 600 rpm. The epoxy was then allowed to cool to room temperature and hardener was added using an epoxy-to-hardener weight ratio of 100:27. Air release agent (BYK A-501) was also added using an air release agent-to-total mixture weight ratio of 0.001:1. The mixture was then stirred by hand for 5 minutes and degassed at room temperature for 20 minutes in a vacuum chamber.

Unidirectional glass fiber sheets (Fibre Glast Saertex) were impregnated with the degassed CNF-modified epoxy using hand lay-up to produce two 10" \(\times\) 10" laminates with stacking sequences of \([0/90]_{\text{s}}\) and thicknesses of 3 mm. We will refer to these as laminates 1 and 2. Both laminates were vacuum bagged and cured in an oven for 5 hours at 60 $^{\circ}$C. After curing, two square specimens measuring 3.25" \(\times\) 3.25" were cut from each laminate. using a water-cooled tile saw. This resulted in a total of four plate-like specimens, one of which is used for later through-hole tests and three of which are used for later impact tests. Additionally, ten smaller specimens measuring 0.25" \(\times\) 0.25" were cut  out of the remaining material from the 10" \(\times\) 10" laminates for the purpose of measuring the average in-plane and through-thickness conductivity of each plate.

\section{Experimental Procedures}

\subsection{Electrical Testing}
\begin{figure}[h!]
	\centering
	\includegraphics[width=0.75\textwidth]{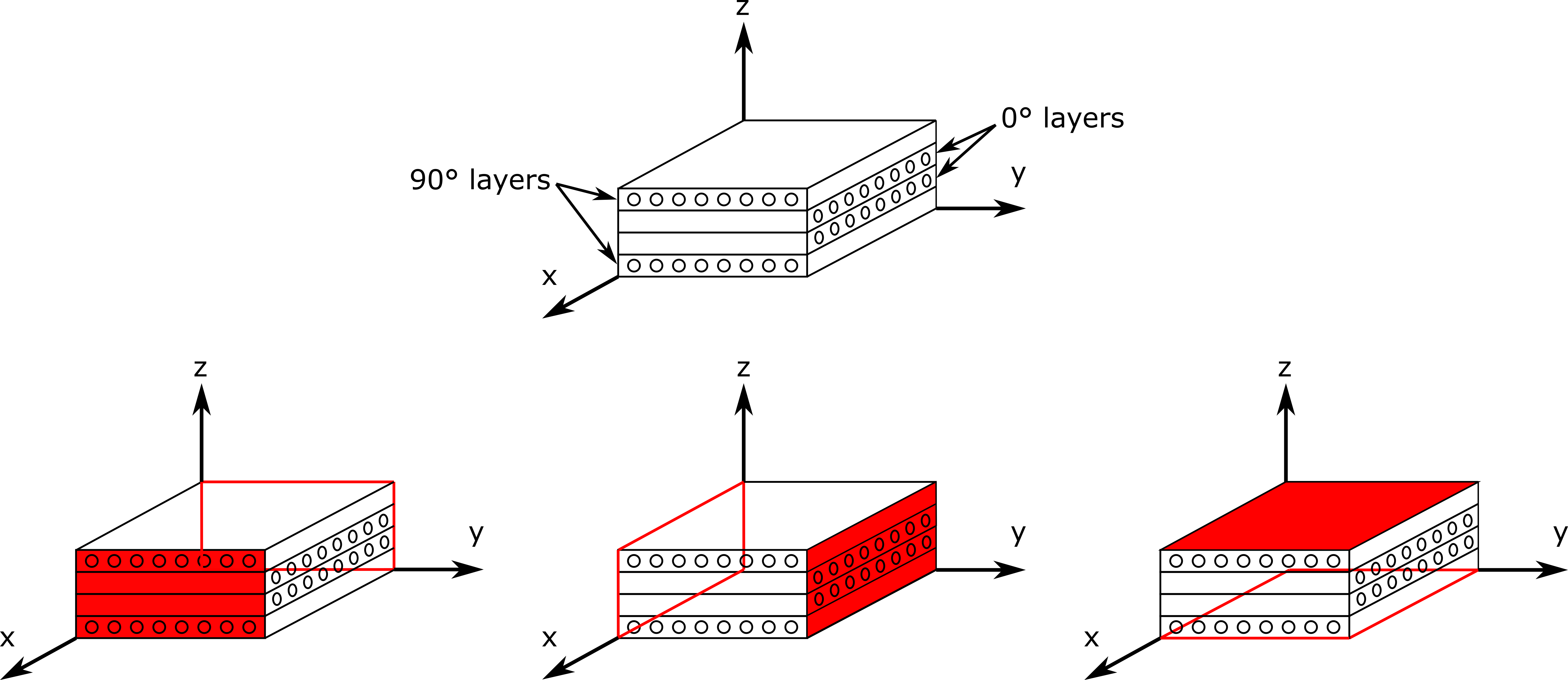}
	\caption{Illustration of conductivity measurement scheme and principal coordinate system. The ellipses represent the fiber cross-sections. Top: Schematic of a sub-section of the laminate showing the principal coordinate system and the layer orientations. Bottom: Conductivity measurement in the \(x\)-(left), \(y\)-(middle), and \(z\)-directions (right). The filled red rectangles represent electrodes on the front faces and the outlined red rectangles represent electrodes on the back faces.}
	\label{fig-cond-meas}
\end{figure}

The conductivities of CNF/GFRP were measured to estimate the baseline conductivity for EIT and also for the damage shaping algorithm. For this, conductivity measurements were collected from the ten 0.25" \(\times\) 0.25" specimens cut from each laminate. The conductivities of each square were measured in the principal (\(x\), \(y\), \(z\)) coordinate system, illustrated in Figure \ref{fig-cond-meas}. To measure the conductivity in the \(x\)-direction, electrodes were applied by painting colloidal silver patches on the faces of the square with normal vectors pointing in the \(\pm x\)-directions. The resistance, \(R\), between the electrodes was measured using a multi-meter and the conductivity was calculated as \(\sigma_x = l_x/RA_e\), where \(l_x\) is the length of the square in the \(x\)-direction and \(A_e\) is the area of an electrode. The conductivities in the \(y\)- and \(z\)-directions were measured similarly. The measurement scheme is illustrated in Figure \ref{fig-cond-meas} and the mean conductivities measured from both laminates are listed in Table \ref{tab-cond}.

\begin{table}
\centering
\caption{Mean and standard deviations of conductivities in the \(x\)-, \(y\)-, and \(z\)-directions measured from laminates 1 and 2.}
\begin{tabular}{cccc}
\toprule
Laminate & \(\sigma_x\) [S/m] & \(\sigma_y\) [S/m] & \(\sigma_z\) [S/m] \\
\midrule
1 & \(0.024\pm0.003\) & \(0.023\pm0.004\) & \((3.5\pm1.8)\times10^{-4}\) \\
2 & \(0.054\pm0.003\) & \(0.057\pm0.004\) & \((5.6\pm2.2)\times10^{-4}\) \\
\bottomrule
\end{tabular}
\label{tab-cond}
\end{table}

\subsection{Through-Hole and Impact Testing}
\begin{figure}
    \centering
    \subfloat{{\includegraphics[width=0.30\textwidth]{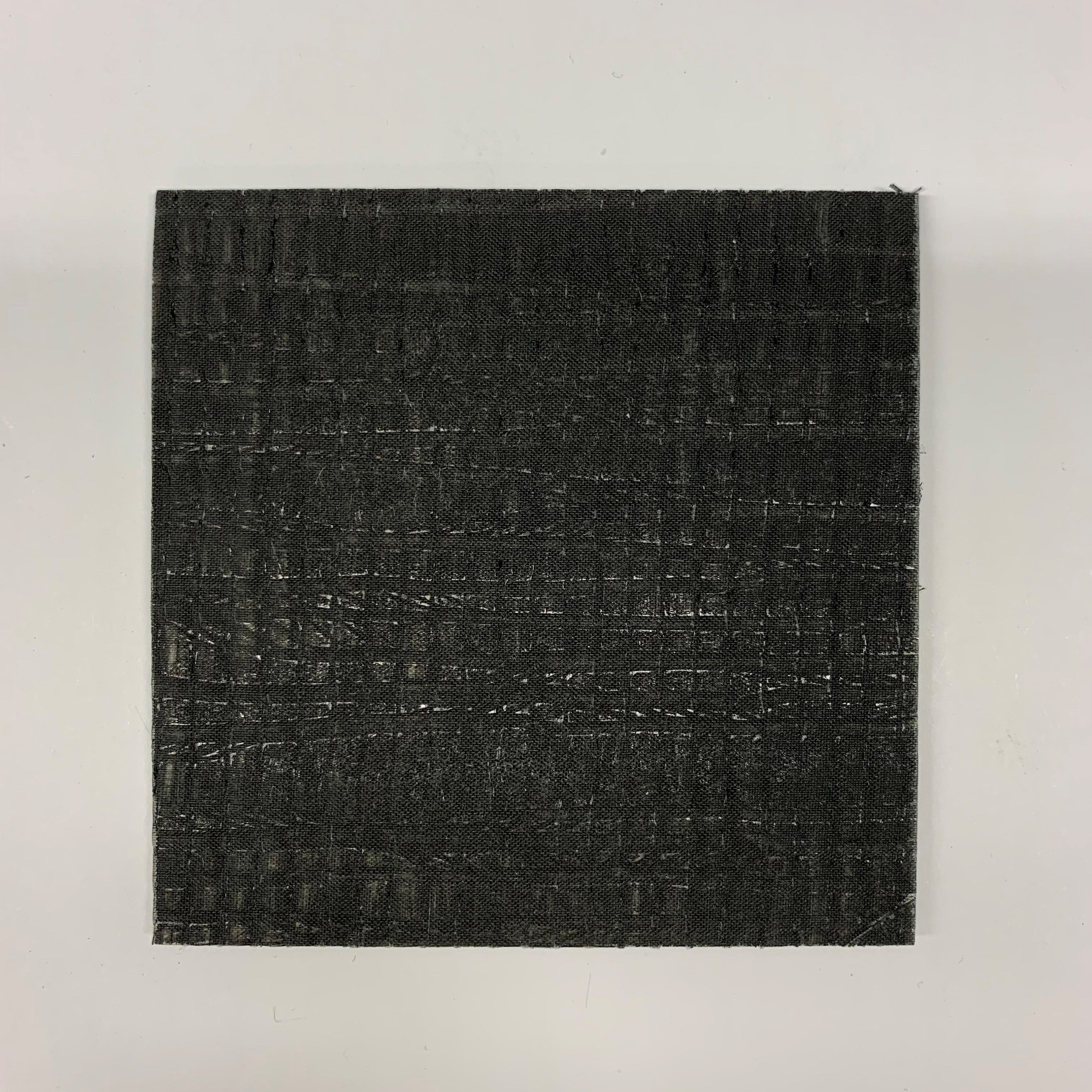} }}%
    \qquad
    \subfloat{{\includegraphics[width=0.53\textwidth]{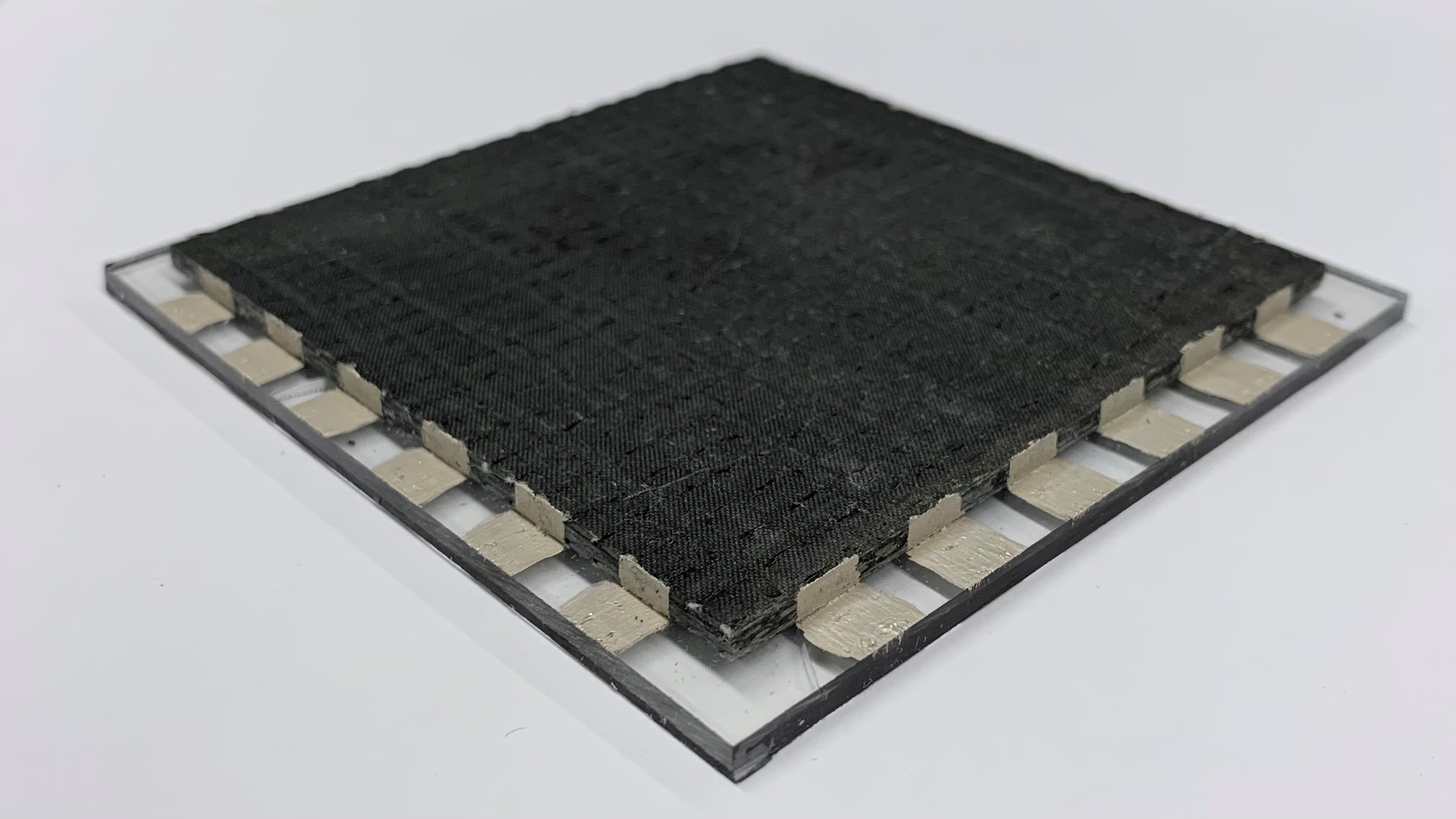} }}%
    \caption{Representative CNF/GFRP square laminate used for through-hole and impact testing. Left: Top-down view of laminate without electrodes. Right: Laminate with electrodes painted and adhered to acrylic base.}%
    \label{fig-specimen}%
\end{figure}

The four 3.25" \(\times\) 3.25" specimens were used for through-hole and impact testing as follows. For the two specimens cut from laminate 1, one was used for through-hole testing and the other was impacted with 25 J. For the two specimens cut from laminate 2, one was impacted with 23 J and the other was impacted with 28 J. Note, however, that all specimens tested were made using the same procedure described above and used the same CNF weight fraction. In order to collect EIT data, electrodes were applied to each specimen by painting evenly spaced colloidal silver patches on each edge. Each specimen was then adhered to an acrylic base and additional colloidal silver patches were painted on the acrylic to act as extended electrode tabs. A representative specimen with electrodes painted is shown in Figure \ref{fig-specimen}. Each specimen was connected to a current source (Keithley 6221) and a DAQ system (National Instruments PXIe-6368) to measure the electrode voltages. 

For the through-hole testing specimen, a current magnitude of 0.25 mA was used. One set of voltages was collected from the specimen in its undamaged state. Next, three circular holes of radii 1.19 mm, 2.38 mm, and 3.18 mm were successively drilled and voltages were collected after drilling each new hole.

For the three impact testing specimens, the energies (23 J, 25 J, and 28 J) were chosen deliberately to induce a measurable delamination without completely perforating the specimens. A current magnitude of 0.25 mA was used for the 23 J and 28 J impact specimens while a current magnitude of 0.1 mA was used for the 25 J impact specimen. One set of voltages was collected from each specimen in its undamaged state. The specimens were then mounted on a 6" \(\times\) 4" aluminum plate with a centrally located 2" \(\times\) 2" window and impacted in a drop tower (CEAST 9340) using a steel hemispherical striking head with a diameter of 15.8 mm. Post-impact voltages were then collected from each specimen. All three specimens were then destructively evaluated in order to determine the true size and shape of the delaminations. This was done by making multiple cuts through each specimen using a water-cooled tile saw and examining the delamination length at the cross-section of each cut using a Zeiss Axioskop 2 MAT microscope. Several images along the length of the delamination at each cross-section were taken and were used to reconstruct the full delamination shape.

\section{Results and Discussion}\label{results_and_discuss}
\subsection{Through-Holes}
\begin{figure}[h]
	\centering
	\includegraphics[width=0.60\textwidth]{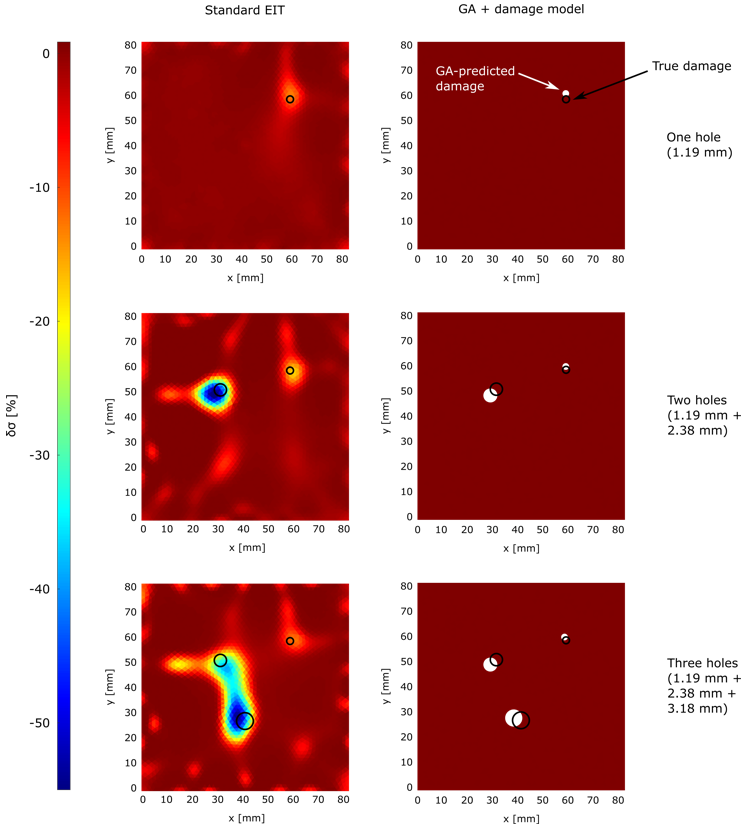}
	\caption{EIT and GA-enabled damage shaping results from through-hole testing. The black circles indicate the true hole sizes and locations. Left column: EIT-imaged conductivity change. We can see that EIT is able to detect the presence of the holes but the exact shape and location are not clear. Right column: GA-enabled damage shaping results. It can be seen that by integrating a realistic damage model with the GA, the shape and size of the holes can be determined much more precisely than standard EIT.}
	\label{fig-eit-ga-holes}
\end{figure}

The mean conductivities measured from laminate 1, shown in Table \ref{tab-cond}, were used as the initial estimate for the baseline conductivity of the through-hole testing specimen. The initial estimate was then iteratively updated to find the optimal baseline conductivity that minimized the difference between the pre-damage voltages and the forward operator predicted voltages. Using this approach, the optimum in-plane baseline conductivities were found to be \(\sigma_{x0}\) = \(\sigma_{y0}\) = 0.025 S/m. EIT was then performed using this baseline on a mesh consisting of 5,670 linear triangular elements and the EIT-imaged conductivity changes are shown in the left column in Figure \ref{fig-eit-ga-holes}. It can be seen from Figure \ref{fig-eit-ga-holes} that EIT is able to successfully detect the presence of the holes. However, it provides little-to-no information about the precise shape and size of each hole. This information must be indirectly inferred from the magnitude of the observed conductivity change. That is, a larger conductivity change corresponds to a bigger hole. 

In order to reconstruct the precise hole size using the damage shaping methodology, a geometric model for through holes must be integrated with the GA. In this case the model consisted of a circular hole inside the FE mesh and the damage shape parameter was specified as \(s_{GA}\) = \([x_c, y_c, r]\), where \(x_c\) and \(y_c\) are the \(x\)- and \(y\)-coordinates of the center of the hole, respectively, and \(r\) is the radius of the hole. The GA initiates the search by generating a population between the bounds \(s_{\text{min}}\) and \(s_{\text{max}}\), which are the lower and upper bounds, respectively, on the location and size of the hole. These are specified based on the size and location of the artifacts observed in the EIT images. For example, for the case of one hole, \(s_{\text{min}}\) was specified as [55, 55, 0.5] mm and \(s_{\text{max}}\) was [62, 62, 5] mm. An adaptive meshing algorithm \cite{engwirda2014locally} was used to integrate this geometric model with the GA. Additionally, a population size of 50, a BSA of 99\%, and a maximum of 30 generations were used for the GA. The results are shown in the right column in Figure \ref{fig-eit-ga-holes}. 

In the first case, the GA is able to reconstruct a hole with a radius of 1.33 mm. In the second case, the GA is able to reconstruct one hole with a radius of 1.34 mm and a second hole with a radius of 2.79 mm. And in the third case, the GA is able to reconstruct three holes with radii of 1.34 mm, 2.79 mm, and 3.43 mm. We can see that these results are considerably more accurate and provide much more precise information about the damage shape and size than standard EIT, which often suffers from background noise and stray artifacts. The average percent errors in the GA predicted hole sizes and locations are listed in Table \ref{tab-hole-error}.

\begin{table}[h]
\centering
\caption{Average percent error in GA-predicted solutions relative to actual hole sizes and locations. Here, \(e_{x_c}\) and \(e_{y_c}\) are the average percent errors in the \(x\)- and \(y\)-coordinates of the center of the hole and \(e_r\) is the average percent error in the hole radius.}
\begin{tabular}{cccc}
\toprule
Hole number & \(e_{x_c}\) [\%] & \(e_{y_c}\) [\%] & \(e_{r}\) [\%] \\
\midrule
1 & 1.67 & 3.33 & 11.76 \\
2 & 3.33 & 5.77 & 17.22 \\
3 & 9.52 & 3.70 & 7.86 \\
\bottomrule
\end{tabular}
\label{tab-hole-error}
\end{table}

\begin{figure}[h]
	\centering
	\includegraphics[width=0.75\textwidth]{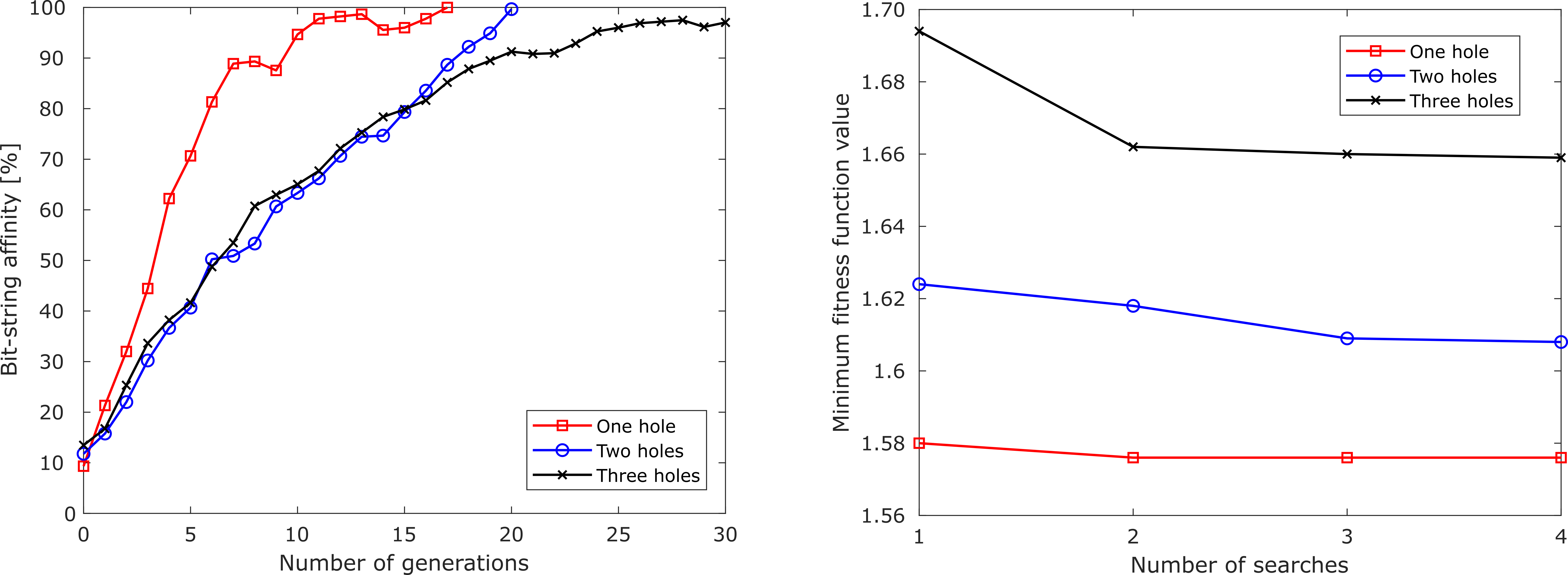}
	\caption{Convergence plots for through-hole reconstruction. Left: BSA convergence for first search case of each case. Right: Fitness function convergence for all searches of each case.}
	\label{fig-holes-conv}
\end{figure}

The BSA convergence for the first search of each case and the fitness function convergence for all searches of each case are shown in Figure \ref{fig-holes-conv}. For one hole, we observe good convergence with one search. For two holes, we require a higher number of generations to reach the BSA stopping criterion. In this case, successive searches slightly improve the solution. For three holes, the maximum number of generations is reached before the BSA stopping criterion is met. In this case, successive searches significantly improve the solution. This is because as the number of holes increases, the number of variables in \(s_{GA}\) also increases and the GA requires more computations to achieve a genetically similar population. 

\subsection{Delaminations}
\begin{figure}[h!]
	\centering
	\includegraphics[width=0.75\textwidth]{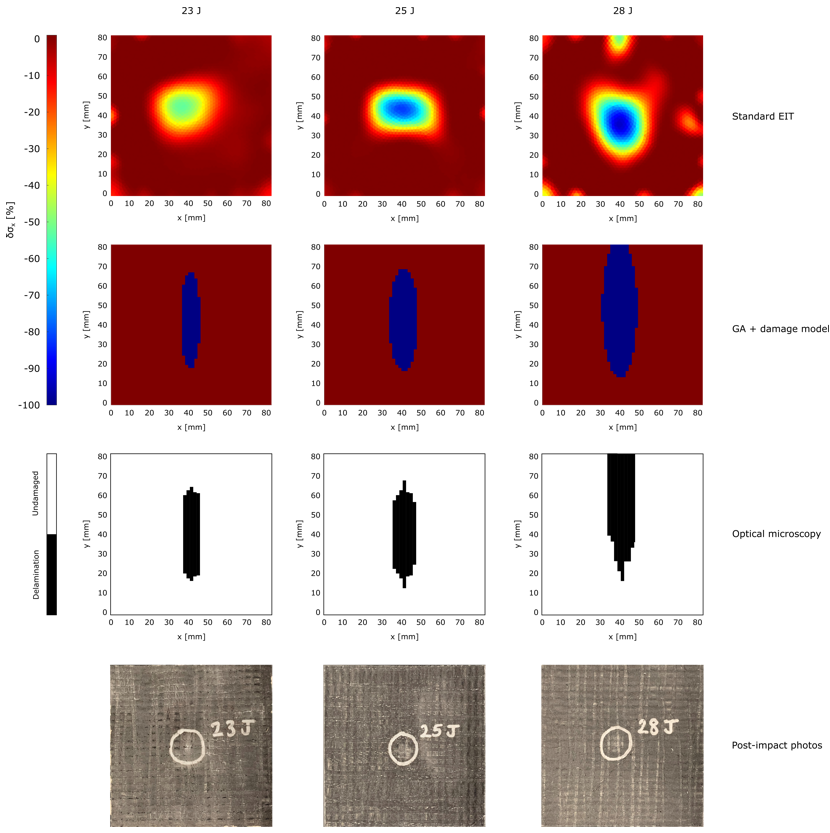}
	\caption{EIT, delamination shaping, and optical microscopy results from impact testing specimens. First (top) row: EIT-imaged conductivity change. The magnitude of the conductivity change increases as the impact energy increases. Second row: GA-enabled delamination shaping results. Third row: Actual delamination shapes reconstructed using optical microscopy. Bottom row: Photos of post-impacted specimens. The silver circles indicate the actual impact locations. Note that damage caused by the impacts is barely visible on the surface of the specimens.}
	\label{fig-eit-ga-delams}
\end{figure}

The optimum baseline conductivities of the impact testing specimens were estimated using a similar approach as the through-hole testing specimen. The conductivities of the 23 J and 28 J impact specimens (from laminate 2) were found to be \(\sigma_{x0}\) = \(\sigma_{y0}\) = 0.055 S/m, and \(\sigma_{z0} = 5\times10^{-4}\) S/m, and the conductivities of the 25 J impact specimen (from laminate 1) were found to be \(\sigma_{x0}\) = \(\sigma_{y0}\) = 0.02 S/m, and \(\sigma_{z0} = 3\times10^{-4}\) S/m. EIT was performed on a mesh of 5,670 linear triangular elements and the conductivity change results from each impact are shown in the top row in Figure \ref{fig-eit-ga-delams}. Note that EIT endeavored to reconstruct the in-plane conductivity, \(\sigma_{x0}\) = \(\sigma_{y0}\) = \(\sigma\), which is isotropic -- even though each lamina is mildly electrically anisotropic, the symmetric layup renders the in-plane conductivity of the entire laminate as very nearly isotropic. We can immediately observe that EIT is able to detect each of the impacts and as the impact energy increases the intensity of the conductivity change at the impact location also increases. This indicates that increasing the impact energy is increasing the amount of damage within the laminate. However, from the EIT image alone, the size and shape of the induced delamination cannot be obviously determined.

\begin{figure}
	\centering
	\includegraphics[width=1.0\textwidth]{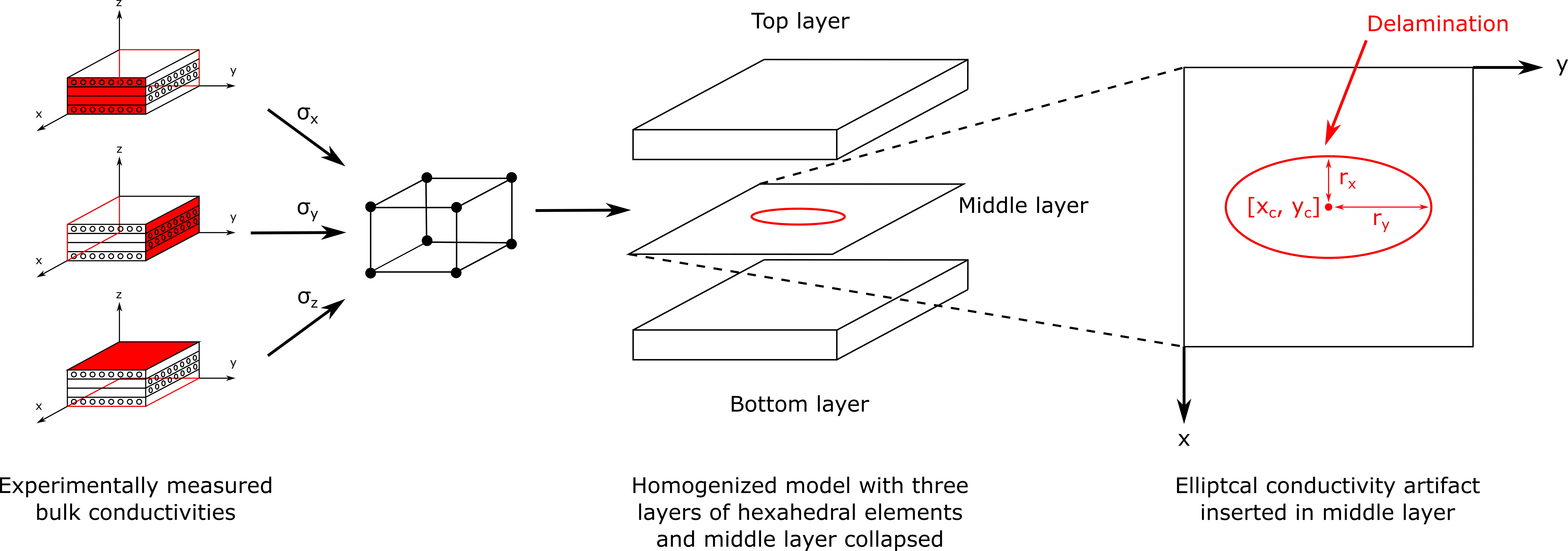}
	\caption{Schematic of delamination model integrated with the GA. The experimentally measured bulk conductivities are used to build a homogenized FE model with three layers of hexahedral elements. The middle layer is then collapsed to infinitesimal thickness and the GA is used to generate elliptical conductivity artifacts within this layer of elements.}
	\label{fig-delam-model}
\end{figure}

To compute the delamination shape and size, the GA was integrated with a geometric model for delaminations as follows. FE models of each laminate were constructed using \(\sigma_{x0}\), \(\sigma_{y0}\), and \(\sigma_{z0}\). These models represent homogenized or `effective' laminates that have the same bulk electrical behavior as the physical laminate. Each FE model consisted of one layer of 2,700 linear quadrilateral elements sandwiched between two layers of 2,700 linear hexahedral elements. The GA was then used to generate elliptical conductivity artifacts described by the shape parameter \(s_{GA}\) = \([x_c, y_c, r_x, r_y]\) inside the middle layer. A conductivity of \(\sigma_x\) = \(\sigma_y\) = \(\sigma_z\) = \(\sigma_d\) = \(1\times10^{-6}\) S/m was assigned to the elements within the elliptical artifact. This concept is based on the fact that a delamination can be modeled as a failure at the interface between two layers of the laminate. In this case, these are the two layers of hexahedral elements that represent the homogenized composite. The individual layers of the physical laminate were not explicitly modeled for the inversion. These were not modeled because the EIT electrodes spanned the thickness of the laminate thereby making it impossible to extract through-thickness information. As such, the GA-produced delamination was inserted at the mid-plane of the homogenized models. A schematic of the delamination modeling procedure is shown in Figure \ref{fig-delam-model}. For the GA, a population size of 20, a BSA of 65\%, and a maximum of 20 generations were used. 

\begin{figure}
	\centering
	\includegraphics[width=0.45\textwidth]{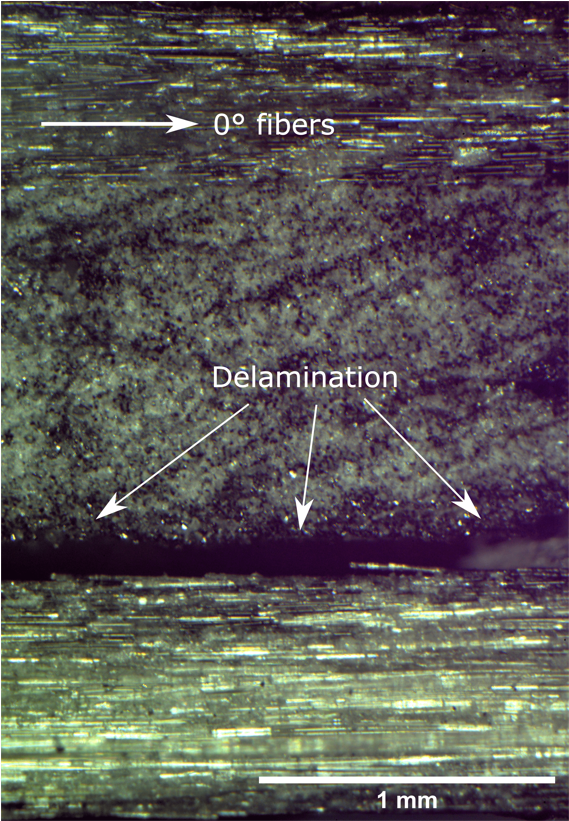}
	\caption{Optical microscopy image of cross-section of 23 J impact specimen. A delamination is clearly visible between the third and fourth layers, as indicated by the white arrows. The fibers in the \(90 ^{\circ}\) layers are pointing out of the page.}
	\label{fig-micro-img}
\end{figure}

\begin{figure}
	\centering
	\includegraphics[width=1.0\textwidth]{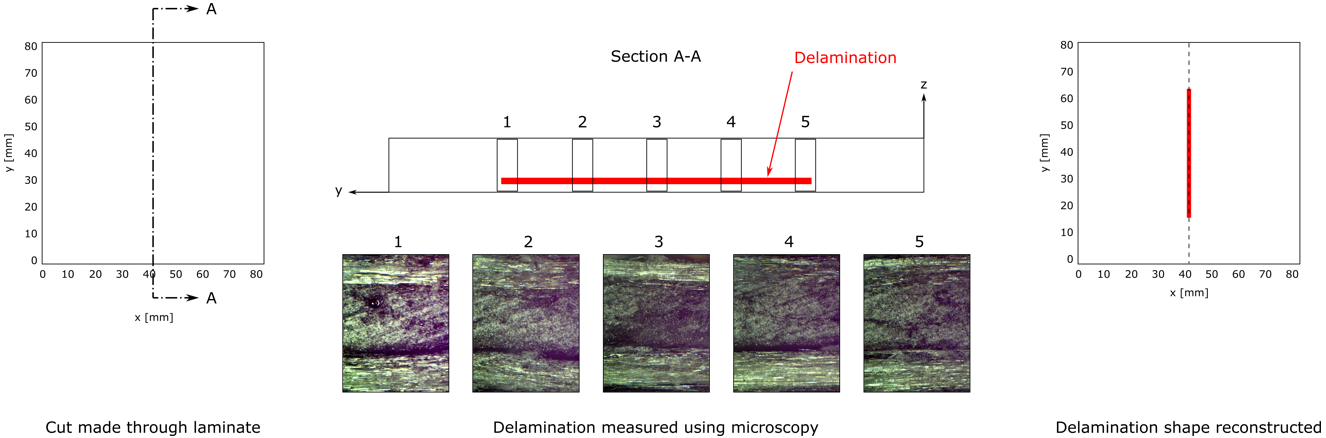}
	\caption{Illustration of actual delamination shape reconstruction using optical microscopy. Left: A cut is made through the laminate to expose the cross-section. Middle: Optical microscopy is used to measure the delamination length at the cross-section. Right: The cut thickness and delamination length are used to piece together the delamination. This procedure is repeated until the full delamination shape is obtained.}
	\label{fig-micro-slices}
\end{figure}

As described earlier, optical microscopy was used to destructively evaluate the specimens. Figure \ref{fig-micro-img} shows an optical micrograph of the cross-section of the 23 J laminate. A delamination is clearly visible between the third (\(0 ^{\circ}\)) and fourth (\(90 ^{\circ}\)) layers. In fact, it was observed that the delamination always occurred between the third and fourth layers for each impact. The full shape of the delamination was then reconstructed as follows. For each cut made through a specimen, the delamination length was recorded and images were taken at regular intervals along the cross-section. A new cut was then made, the delamination length recorded again, and more images taken. In other words, the delamination length was measured with the aid of an optical microscope for each slice. This was repeated until a delamination was no longer observed using the microscope. The recorded delamination lengths and cut thicknesses were then used to reconstruct the in-plane shape of the delamination. This process is illustrated for a single cut in Figure \ref{fig-micro-slices}. 

The GA-generated delamination shapes along with the actual delamination shapes reconstructed using optical microscopy are shown in the second and third rows, respectively, in Figure \ref{fig-eit-ga-delams}. We can see immediately that the GA-generated solutions agree very well with the actual delamination shapes. As the impact energy increases, the size of the delamination also increases, which is what we observe from both the GA results and optical microscopy. There are nonetheless some differences between the GA-generated solutions and the actual delamination shapes. This may be attributed to the fact that an impact does not cause a pure delamination but a combination of delamination, matrix cracking, and fiber breakage. However, the proposed model does not account for these additional damage modes. As such, the algorithm tries to compensate for the additional damage by increasing the size of the predicted delamination. Regardless, we can see from Figure \ref{fig-eit-ga-delams} that the GA-predicted solutions are a significant improvement over standard EIT in terms of damage shape.


\begin{figure}[h!]
	\centering
	\includegraphics[width=0.75\textwidth]{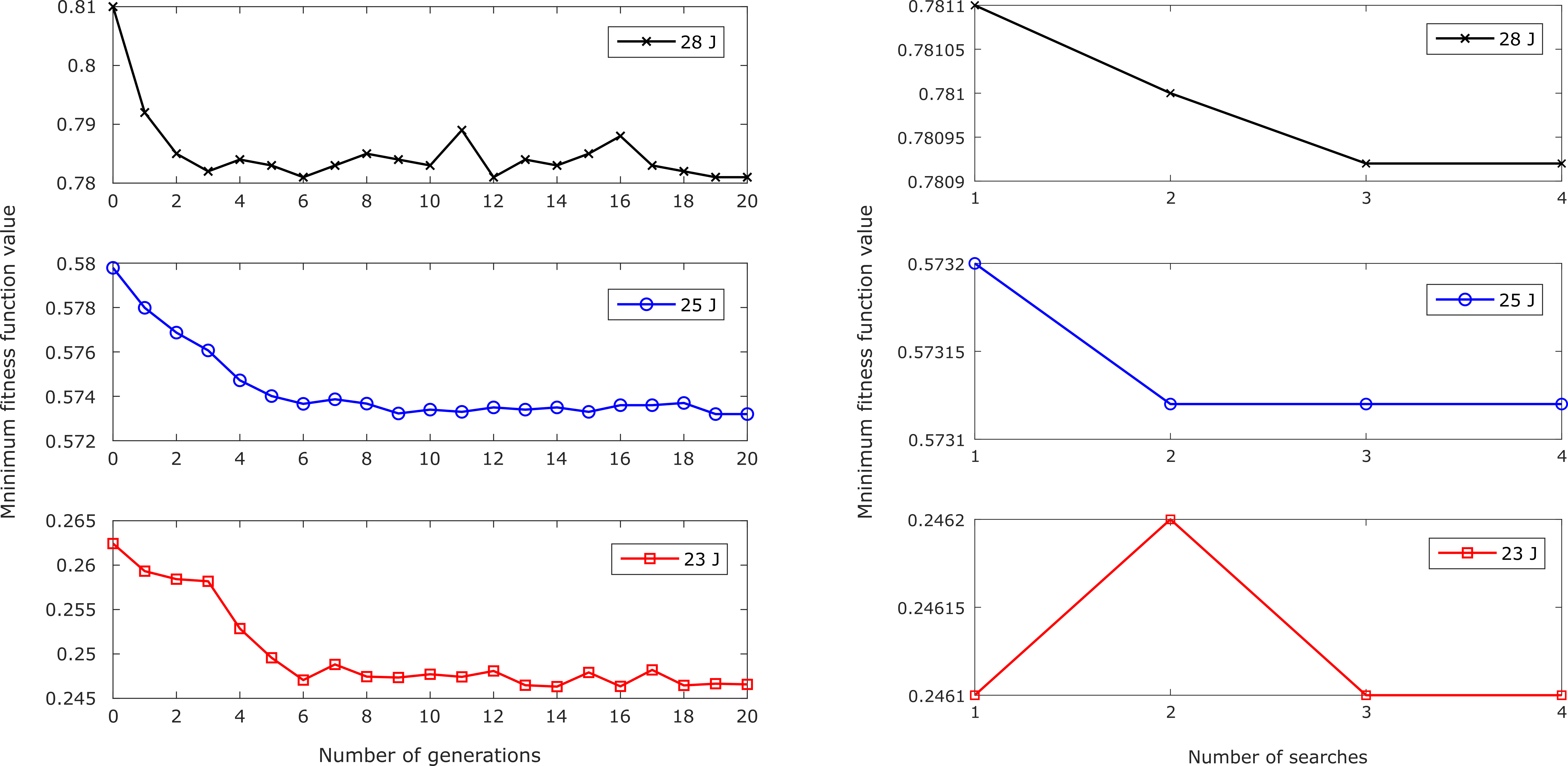}
	\caption{Fitness function convergence plots for GA-enabled delamination shaping algorithm. Left: Fitness function convergence for first search of each impact case. Right: Fitness function convergence for all searches of each case.}
	\label{fig-delam-conv}
\end{figure}

Lastly, we examine the convergence of the delamination shaping algorithm. The fitness function convergence for the first search and the minimum fitness function value for all searches are shown in Figure \ref{fig-delam-conv}. We can see that the fitness function attains its minimum value for a relatively small number of generations. Additionally, the change in the minimum fitness function value between successive searches is not significant. This indicates that successive searches may not necessarily produce better quality solutions for the delamination shaping problem.

\section{Summary and Conclusions}
In this article we have presented a new technique for more accurately determining the shape and size of damage in self-sensing composites via EIT. Our technique uses a GA to inversely determine the shape of particular damage modes from EIT-imaged conductivity changes and boundary voltages. This is a significant advancement to state-of-the-art conductivity-based SHM which seeks only to spatially localize damage and does not provide any information about the underlying mechanics. The technique proposed here addresses this limitation and can transform SHM and NDE in self-sensing structures from mere damage detection and localization to much more robust damage characterization.

We began by manufacturing self-sensing CNF-modified GFRP laminates. The laminates were damaged by drilling multiple through-holes and impacting with three different energies to cause a delamination. EIT was performed on each laminate to image the damage-induced conductivity change. Two physics-based geometric models for damage were then developed---one for through-holes and one for delaminations---and each was integrated with a GA. The GA-integrated damage shaping algorithms were then used to reconstruct the shape and size of the holes and delaminations based on the observed conductivity changes and boundary voltages. The inversely reconstructed through-holes showed good agreement with the actual hole sizes. Furthermore, the algorithm was able to accurately reconstruct multiple holes. The delamination shaping results also showed good agreement with the actual delamination shapes determined destructively using optical microscopy.
The work presented here has potential to improve the safety of self-sensing structures in safety-critical applications by providing precise information about damage condition. However, some limitations of this approach must be recognized and addressed in future work. First, the use of a GA adds significant computational burden to EIT. We anticipate that this can be overcome by using more efficient high-performance computing architecture. Second, although this approach seems to show much potential, this exploratory study was limited to through-holes and delaminations. Additional damage models should be developed to capture more realistic and complex damage scenarios. Appropriate damage models will have to developed for these cases too. Finally, machine learning and artificial neural networks have shown promise in solving damage inversion and categorization problems \cite{lin2020realtime}. These should also be explored in future work.

\section*{Acknowledgements}
The authors gratefully thank Professor WA Crossley at Purdue University for the use of his GA, and Professor W Chen and graduate research assistant Nesredin Kedir at the Impact Science Lab at Purdue University for their assistance with optical microscopy. The authors also thank the Purdue University Graduate School for providing financial support through the Bilsland Dissertation Fellowship. 

 \bibliographystyle{elsarticle-num} 
 \bibliography{cas-refs}





\end{document}